\documentclass[aps,prl,twocolumn,superscriptaddress,showpacs]{revtex4-1}
\usepackage[active]{srcltx}
\usepackage{graphicx}
\usepackage{amsmath,amssymb}
\begin{document}

%\preprint{}

\title{Two-dimensional magnetism in \mbox{$\kappa$-(BEDT-TTF)$_2$Cu[N(CN)$_2$]Cl}, a spin-1/2 Heisenberg antiferromagnet with Dzyaloshinskii-Moriya interaction }


\author{\'Agnes~Antal}
\affiliation{Budapest University of Technology and Economics, Department of Physics and Condensed Matter Research Group of the Hungarian Academy of Sciences, P.O. Box 91, H-1521 Budapest, Hungary}

\author{Titusz~Feh\'er}
\affiliation{Budapest University of Technology and Economics, Department of Physics and Condensed Matter Research Group of the Hungarian Academy of Sciences, P.O. Box 91, H-1521 Budapest, Hungary}

\author{B\'alint~N\'afr\'adi}
\affiliation{Budapest University of Technology and Economics, Department of Physics and Condensed Matter Research Group of the Hungarian Academy of Sciences, P.O. Box 91, H-1521 Budapest, Hungary}
\affiliation{Institute of Physics of Complex Matter, FBS, Swiss Federal Institute of Technology (EPFL), CH-1015 Lausanne, Switzerland}

\author{L\'aszl\'o~Forr\'o}
\affiliation{Institute of Physics of Complex Matter, FBS, Swiss Federal Institute of Technology (EPFL), CH-1015 Lausanne, Switzerland}

\author{Andr\'as~J\'anossy}
\email{atj@szfki.hu}
%\thanks{}
\affiliation{Budapest University of Technology and Economics, Department of Physics and Condensed Matter Research Group of the Hungarian Academy of Sciences, P.O. Box 91, H-1521 Budapest, Hungary}


\date{\today}

\begin{abstract}
Field-induced antiferromagnetic (AF) fluctuations and magnetization are
observed above the (zero-field) ordering temperature, $T_N=23\rm\,K$ by
electron spin resonance in
\mbox{$\kappa$-(BEDT-TTF)$_2$Cu[N(CN)$_2$]Cl}, a quasi two-dimensional
antiferromagnet with a large isotropic Heisenberg exchange
interaction. The Dzyaloshinskii-Moriya (DM) interaction is the main
source of anisotropy, the exchange anisotropy and the
interlayer coupling are very weak. The AF magnetization is induced by
magnetic fields perpendicular to the DM vector; parallel fields have no
effect. The different orientation of the DM vectors and the $g$ factor
tensors in adjacent layers allows the distinction between interlayer and
intralayer correlations. Magnetic fields induce the AF magnetization
independently in adjacent layers. We suggest that the phase transition
temperature, $T_N$ is determined by intralayer interactions alone.
\end{abstract}

%\keywords{}

\maketitle

%\section{}
%\subsection{}
%\subsubsection{}

One-dimensional (1D) magnets, i.e., isolated chains of magnetic atoms or
molecules have no ordering phase transitions while 3D magnets order at
finite temperatures, $T_N$. Magnetic ordering in two dimensions, i.e.,
in isolated planes, is a delicate question \cite{Mattis2006}. Theory
predicts for most anisotropic two-dimensional (2D) magnets a phase transition
at finite $T_N$ except if ordering breaks a continuous symmetry. The case of
$S=1/2$ anisotropic Heisenberg antiferromagnets is complicated
\cite{Frohlich1977}. Experiments on low dimensional magnets with
negligible interactions between the magnetic chains or planes are rare.
The magnetic-field-induced change in the excitation spectrum of 1D
Heisenberg chains \cite{Dender1997,Zvyagin2005} agrees with theory
\cite{Oshikawa1997,Affleck1999}. Field-induced magnetism has also been
studied in a quasi 2D $S=1$ dimer system \cite{Kodama2005}. The present
experiments are on a quasi 2D, $S=1/2$ magnet where the continuous
symmetry is broken by a very small anisotropy and the applied magnetic
field, $H$.

% Mattis2006 See D.C. Mattis The Theory of Magnetism Made Simple World Scientific Singapore 2006 and references therein
% Shenker1980  S.H. Shenker, J. Tobochnik PRB22 4462 (1980)
% Frohlich1977 J. Frohlich E.H. Lieb PRL38 440 (1977), ??Commun Math. Phys. Vol.60, 233 (1978)
% Frohlich Simon Spencer Commun Math. Phys.50, 79 (1976)
% Mermin1966 PRL 17 1133 1307

\begin{figure}
\includegraphics[angle=0,width=8.2cm] {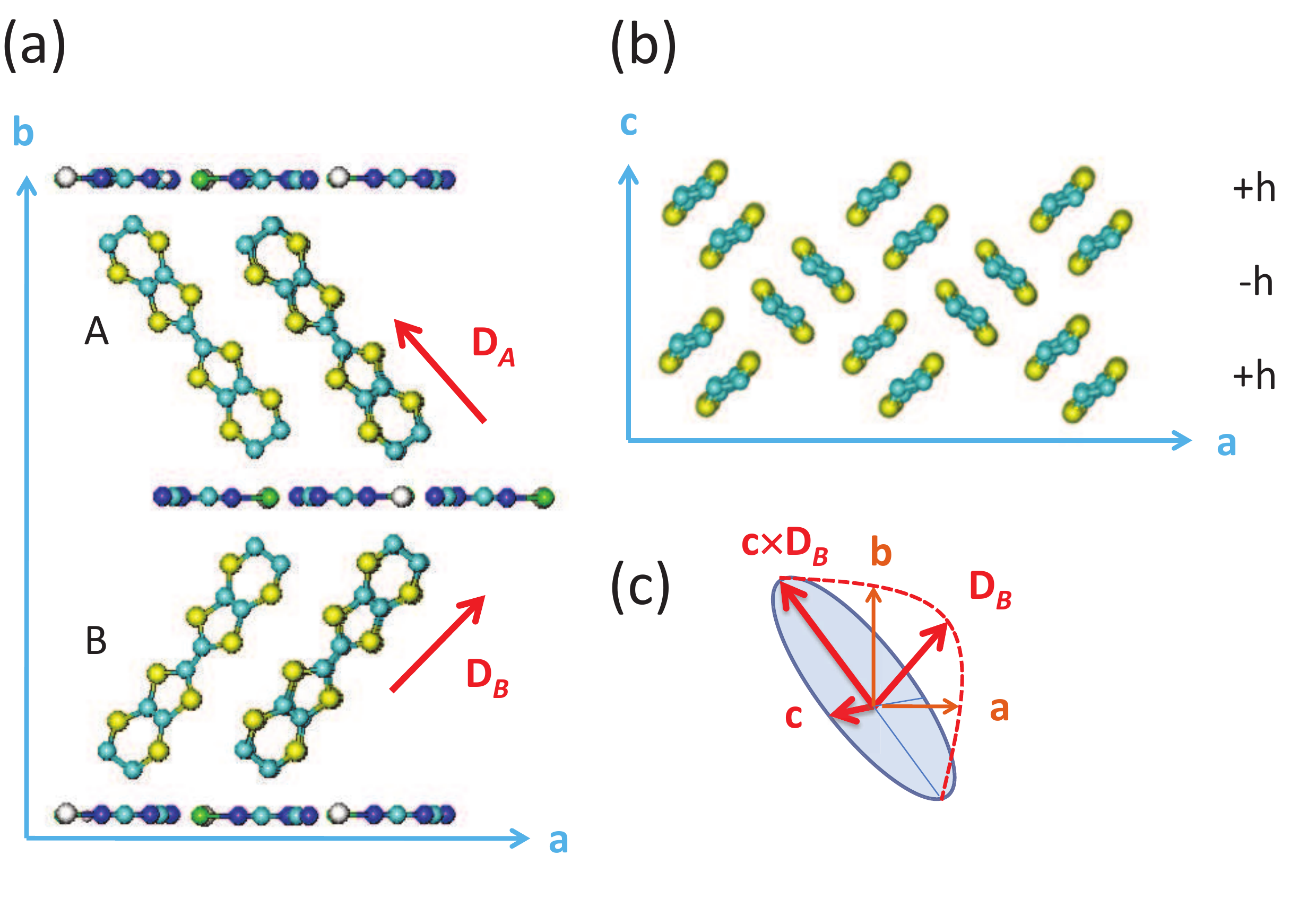}
\caption{(Color online) Structure of \mbox{$\kappa$-ET$_2$-Cl}. (a)
Projection along $\mathbf{c}$ onto the $(\mathbf{a},\mathbf{b})$ plane.
Only two ET molecules per plane are shown for clarity. The
Dzyaloshinskii-Moriya vectors of the $A$ and $B$ layers, $\mathbf{D}_A$
and $\mathbf{D}_B$ are oriented differently in the
$(\mathbf{a},\mathbf{b})$ plane. (b) The 2D magnetic ET layer projected
along the long ET molecular axis onto the $(\mathbf{a},\mathbf{c})$
plane. The effective field, $\mathbf{h}$, due to the DM interaction
alternates along $\mathbf{c}$. (c) Magnetic field orientations of ESR
measurements in layer $B$. The field-induced magnetization in the plane
perpendicular to the DM vector (blue ellipse) shifts and broadens the
ESR.\label{fig:lstruct} }
\end{figure}

\mbox{$\kappa$-(BEDT-TTF)$_2$Cu[N(CN)$_2$]Cl}
(\mbox{$\kappa$-ET$_2$-Cl}) is a layered spin 1/2 Heisenberg
antiferromagnet with a magnetic ordering transition \cite{Welp1993} at
$T_N=23\rm\,K$  measured \cite{Kagawa2008} in $H=0$. The ET molecules are
arranged in a 2D lattice of singly charged dimers. The electronic band
is effectively half filled and the system is on the insulating side of a
nearby metal-insulator Mott transition. The two chemically equivalent
organic ET layers \cite{Williams1990, KINI1990}, $A$ and $B$
(Fig.~\ref{fig:lstruct}) are separated by Cu[N(CN)$_2$]Cl polymer sheets.

The ordered state is well described by two-sublattice antiferromagnetic
layers weakly coupled through the polymeric sheets \cite{Antal2009}. The
measured macroscopic parameter of the in-plane isotropic exchange,
$(J/2)\sum_{i,j}\textbf{S}_i\cdot\textbf{S}_j$ is $\lambda
M_{0}=2J/(g\mu_{B})=450$~T \cite{Welp1993,smith2003}. The magnitude of
the antisymmetric Dzyaloshinskii-Moriya (DM) exchange interaction,
$(1/2)\sum_{i,j}\textbf{D}_{ijl}\cdot(\textbf{S}_i\times\textbf{S}_j)$
is $DM_{0}=2D_{12}/(g\mu_{B})=3.7$~T \cite{smith2004, Antal2009}.
$\mathbf{D}_A$ ($\mathbf{D}_B$) is aligned  $\varphi_{0}=134^{\circ}$
($46^{\circ}$) from $\textbf{a}$ in the (\textbf{a},\textbf{b}) plane
\cite{smith2004}. [The summation is over the four first-neighbor dimers
$i$ and $j$ [Fig.~\ref{fig:lstruct}(b)], $l=A,B$. $M_0$ is the $ T=0 $
sublattice magnetization. The length of the (\textbf{a},\textbf{b})
plane component of the DM interaction, $D_{12}=|D^{ab}_{ijl}|$ is the
same for all pairs]. The continuous rotational symmetry around the DM
vector is nearly perfect in the absence of $H$; the coupling between
planes and the in-plane anisotropy of the exchange are all of the order
of 1~mT \cite{Antal2009}.

The peak temperature $T^{\ast}$ in the nuclear spin relaxation rate,
$1/T_1$, often identified with the onset of the magnetic order
\cite{Miyagawa1995}, increases significantly with $H$ \cite{Hamad2005}.
This is surprising since in usual antiferromagnets with no anisotropic
interactions  the order is little affected by $H$. Hamad \emph{et
al}.~\cite{Hamad2005} explained changes in 1/$T_1$ under pressure near
$T_N$ by the frequency dependence of the spin fluctuation spectrum. In
their theory, spin fluctuations depend on the frustration of the
exchange interactions in the triangular lattice of the ET dimers. The
theory fails, however, to account for the increase in $T^{\ast}$ with
increasing NMR frequency.

Kagawa \emph{et al}.~\cite{Kagawa2008} pointed out that in
\mbox{$\kappa$-ET$_2$-Cl} the DM interaction plays an important role
above $T_N$. In magnetic fields (unless parallel to the DM vector) the
phase transition is replaced by a crossover. In this case the AF
magnetization remains finite above $T_N$ and there is a weak
ferromagnetic moment along $\mathbf{H}$ at all temperatures. They
measured $\mu_s$, the field-induced staggered static magnetic moment per
sublattice site, to temperatures well above $T_N$ by NMR. They explain
the increase of $T^{\ast}$, with  $H$ by a shift to higher temperatures
of the finite peak in the temperature dependence of
$\chi(Q_{AF},\omega=0)$ replacing the divergence in $H=0$ at $T_N$.
Their mean field model describes qualitatively the $H$ dependence of the
static staggered susceptibility, $\chi(Q_{AF},\omega=0)$ but unrealistic
exchange interaction parameters are required for a quantitative
agreement.

In this paper we study the electron spin resonance (ESR) in
\mbox{$\kappa$-ET$_2$-Cl} above $T_N$ in fields up to 15 T. The
field-induced staggered magnetization, $\mathbf{M}_s$ shifts the
resonance field. Fluctuations of $\mathbf{M}_s$ enhance the transverse
spin relaxation rate, $1/T_2$ and increase the ESR linewidth, $\Delta
H=(\gamma T_2)^{{-1}}$. The fluctuations increase rapidly with
increasing $\mathbf{H}$ in the plane perpendicular to $\mathbf{D}$,
while fields parallel to $\mathbf{D}$ have no effect. The difference in
the orientation of the DM vectors and the $g$ factor tensors in adjacent
weakly coupled $A$ and $B$ layers [Fig.~\ref{fig:lstruct}(a)] allow the
distinction between interlayer and intralayer correlations. Above $T_N$,
$\mathbf{H}$ affects the magnetic order in adjacent layers
independently. We suggest that in zero external field $T_N$ depends on
intralayer interactions alone, i.e., it is the magnetic ordering
temperature of the isolated 2D molecular plane. The nature of the order
in the third dimension (i.e., from plane to plane) is uncertain.
Furthermore, we propose that the increase in $T^{\ast}$ with field in
\mbox{$\kappa$-ET$_2$-Cl} arises from a low frequency field-induced gap.

\begin{figure}
\includegraphics[angle=0,width=8.2cm] {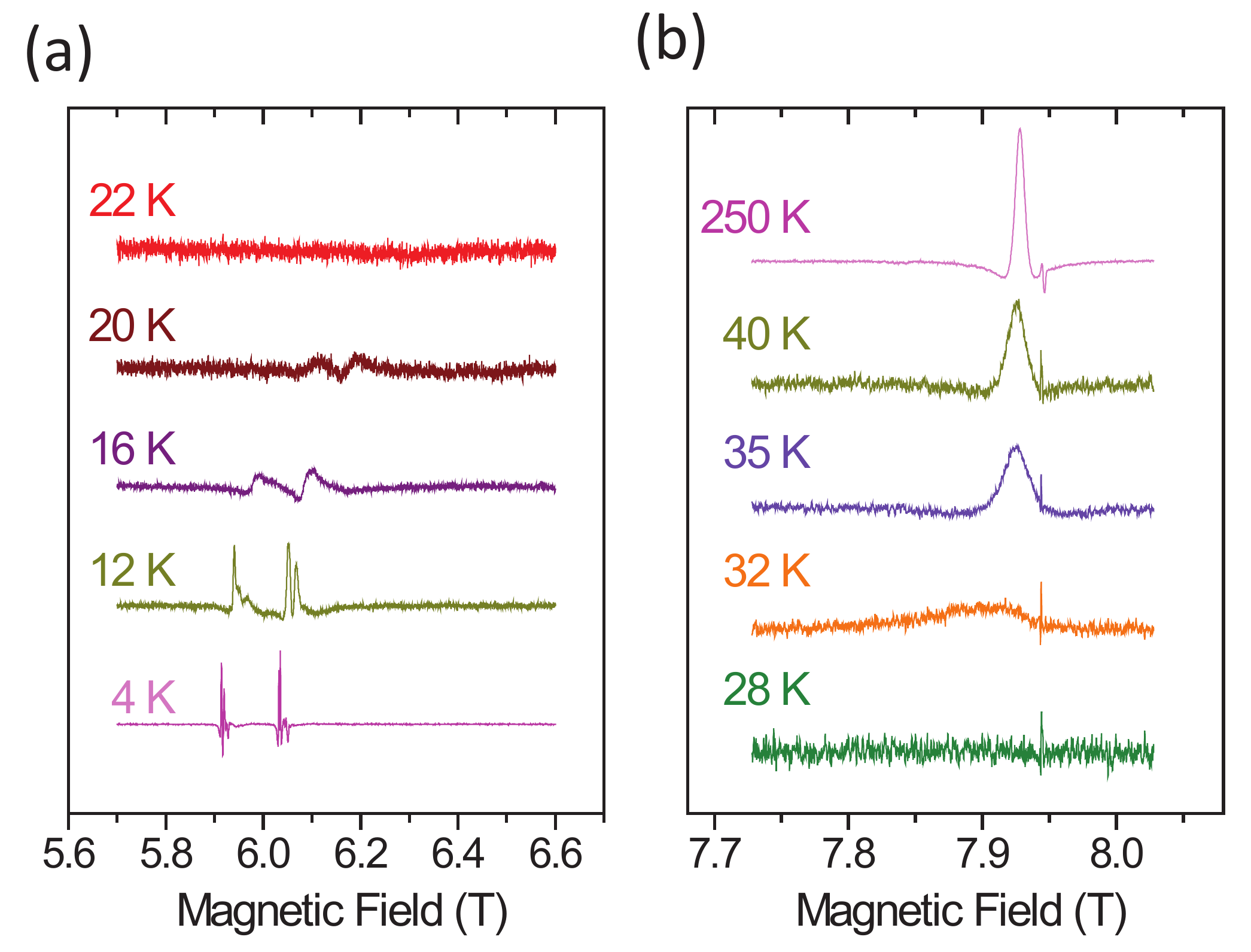}
\caption{(Color online) Temperature dependence of
\mbox{$\kappa$-ET$_2$-Cl} ESR dispersion derivative spectra at 222.4 GHz
frequency and $\mathbf{H}\parallel \mathbf{c}$ magnetic field. The
relative amplitudes at different temperatures are arbitrary. (a) Two
antiferromagnetic resonance modes below $T_N=23\rm\,K$. (b) ESR in the
paramagnetic region. The shift and broadening near $T_N$ is attributed
to the AF magnetization induced in the planes perpendicular to the DM
vectors.\label{fig:cdir} }
\end{figure}

Single crystals of \mbox{$\kappa$-ET$_2$-Cl} were grown by the standard
electrochemical method. Crystal quality was verified by X-ray
diffraction. The ESR spectrometers \cite{Nagy2011,Nafradi2008} operate
at 111.2 and 222.4~GHz at BME \cite{Nagy2011} and at 9, 210 and 420~GHz
at EPFL \cite{Nafradi2008}. The fluctuation spectrum is field dependent
and we often refer to $H$, the approximate ESR magnetic fields instead
of the fixed excitation frequency, $\omega_{+}/2\pi$. We denote the 9.4
GHz ESR data centered at 0.34 T as ``low field'' and the 111.2, 222.4
and 420 GHz data centered near 4, 8 and 15 T as ``high field''. Our low
field spectra in the $\mathbf{b}$ direction agree with the low field ESR
of Yasin \cite{Yasin2008} who measured \mbox{$\kappa$-ET$_2$-Cl} in the
three principal directions. At low fields the ESR of the $A$ and $B$
layers merge into a single line at the average $g$ factor and linewidth.
At high fields there is a single ESR line in the principal directions.
For field angle $\varphi_{ab}\approx45^{\circ}$ in the
$(\mathbf{a},\mathbf{b})$ plane the $A$ and $B$ ESR lines are resolved.
The field direction is very close to $\mathbf{c}\times\mathbf{D}_A$ and
$\mathbf{D}_B$ in the $A$ and $B$ layers, respectively
[Fig.~\ref{fig:lstruct}(c)]. Using the known orientation of the $g$
tensor \cite{NAKAMURA1994}, the lower field ESR line is assigned to the
layer with $\mathbf{H}$ roughly parallel to the long ET molecular axis.

Fig.~\ref{fig:cdir} shows the magnetic resonance spectra from the weak
ferromagnetic ground state to the high temperature paramagnetic state in
8~T magnetic field parallel to $\mathbf{c}$. Near $T_N$, between 22 and
30~K, the resonance broadens beyond observability. The line broadening
and shift depend strongly on the direction and magnitude of the magnetic
field even in the paramagnetic state above $T_N$. In Fig.~\ref{fig:cdir}
$\mathbf{H}$ is perpendicular to both $\mathbf{D}_A$ and $\mathbf{D}_B$
and the shift and broadening are large.

\begin{figure}
\includegraphics[angle=0,width=8.4cm] {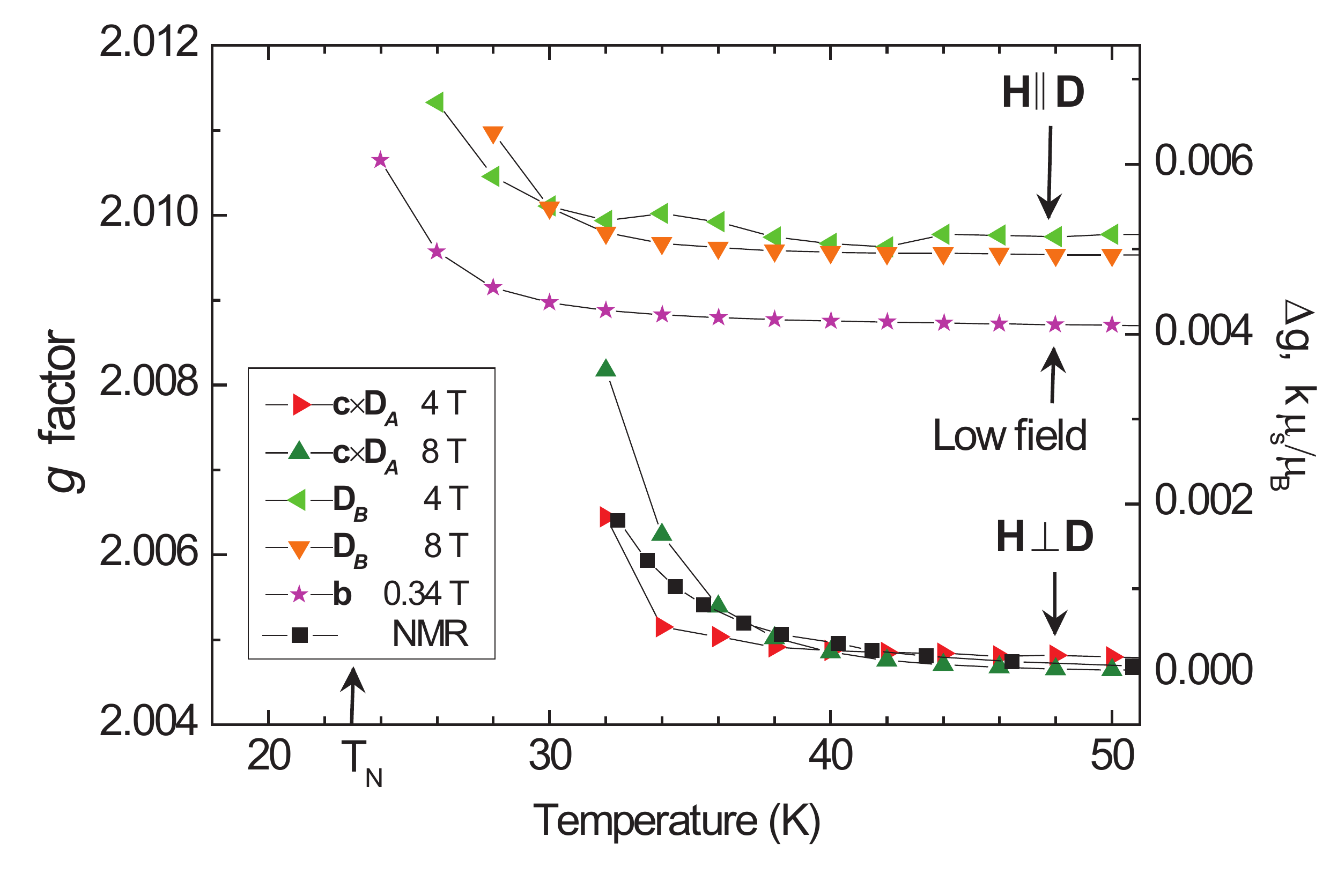}
\caption{(Color online) Temperature dependence of the $g$ factors (left
scale) and $\Delta g$ (right scale) in various fields in
\mbox{$\kappa$-ET$_2$-Cl}. $\blacksquare$: (right scale)
$k\mu_s/\mu_{B}$ from the $^{13}$C Knight shift of
Ref.~\onlinecite{Kagawa2008} in $H=7.4$~T along $\textbf{a}$
($H_\perp=5.2\rm\,T$). $k=g_{0}DM_{0}/(8H_{\perp})$ relates the ESR shift to
the staggered  magnetic moment $\mu_s$ [see Eq.~(\ref{eq3})].
\label{fig:gfactor} }
\end{figure}

The field dependence of the ESR line position, $H_{\mathrm{res}}$, shows
unambiguously that the interlayer coupling is negligible and the
magnetization is induced independently in the $A$ and $B$ layers. The
$g$ factor defined by $\hbar\omega_+=g\mu_BH_{\mathrm{res}}$ is
temperature ($T$) and $\mathbf{H}$ independent above 50~K.
Fig.~\ref{fig:gfactor} displays the $T$ and $\mathbf{H}$ dependence below
50~K. The right scale of Fig.~\ref{fig:gfactor} gives the difference $\Delta
g=g-g_0$ where $g_0$ is the $g$ factor at 50~K for
$\mathbf{H}\parallel\mathbf{c}\times\mathbf{D}$. In $B$ layers where
$\mathbf{H}\parallel\mathbf{D}_B$, $\Delta g$ is independent of $H$. In
contrast, in $A$ layers where $\mathbf{H}\perp\mathbf{D}_A$, $\Delta g$
is anomalously field dependent: as $T\rightarrow T_N$ from above,
$\Delta g$ has an upturn at several degrees higher temperatures in high
fields than in low field. In usual antiferromagnets the upturn of
$\Delta g$ signifies the onset of magnetic correlations. As discussed
later, the field dependence of the $g$ factor is explained by the field
induced staggered magnetization. The $\Delta g$ upturn shifts with field
to higher $T$ only in every second layer. The anomalous field dependent
$g$ shift arises from magnetic correlations within the $A$ layers.
$\Delta g$ is independent of $H$ in layers with
$\mathbf{H}\parallel\mathbf{D}_B$. Even close to $T_N$, interlayer
coupling that would increase the staggered magnetization in $B$ layers
is insignificant.

\begin{figure}
\includegraphics[angle=0,width=8.4cm] {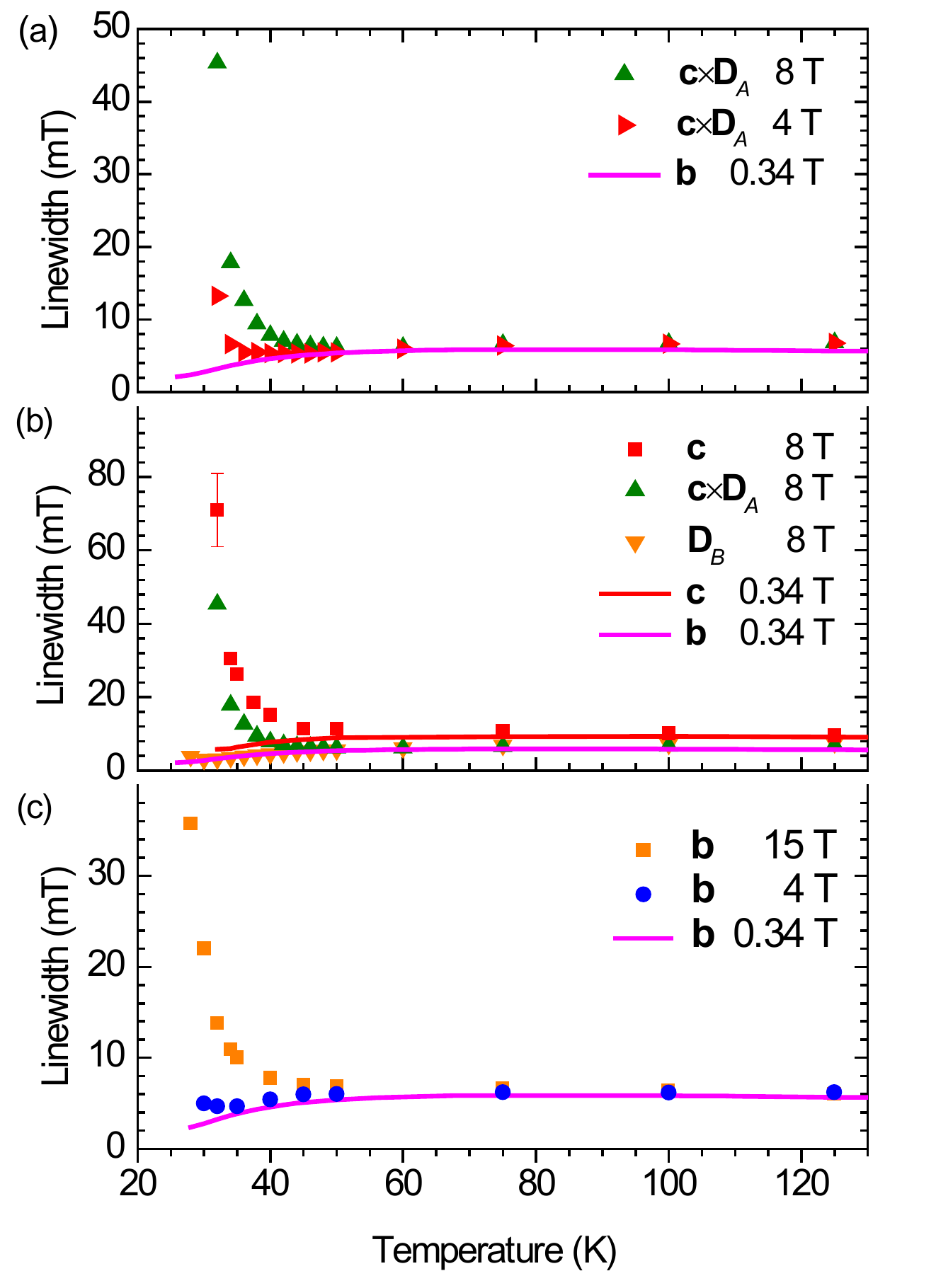}
\caption{(Color online) Frequency and magnetic field direction
dependence of the ESR linewidth, $\Delta H$. (a) The frequency
dependence in the $\mathbf{H}\perp \mathbf{D}$ plane where the
field-induced contribution is largest. (b) Dependence on the direction
of $\mathbf{H}$ with respect to $\mathbf{D}_A$. $\Delta H$ is frequency
independent for $\mathbf{H}\parallel \mathbf{D}$. Data for
$\mathbf{H}\parallel \mathbf{c}$ at 0.34~T are from
Ref.~\onlinecite{Yasin2008}. (c) Frequency dependence of $\Delta H$ for
$\mathbf{H}\parallel \mathbf{b}$.
\label{fig:linewidth} }
\end{figure}

Data on the magnetic-field-induced ESR linewidth are presented in
Fig.~\ref{fig:linewidth}. The half width at half maximum linewidths are
determined from a fit to Lorentzian derivative line shapes. The coupling
between adjacent layers is very small but not strictly zero. For
magnetic fields along $\varphi_{ab} =45^{\circ}$, a detailed analysis of
the lineshape \cite{Antal2011} reveals a small static exchange field
proportional to the homogeneous magnetization below 40~K. At 32~K and
7.5~T the interlayer exchange field is antiferromagnetic and about 1~mT.
In the present work interplane magnetic coupling was taken into account
by fitting the ESR spectra to two Lorentzians with different admixtures
of the absorption and dispersion components.

There are two contributions to the linewidth in the paramagnetic state
below about 50~K: $\Delta H = \Delta H_{0}(T)+\Delta
H_{\mathrm{fl}}(\mathbf{H},T)$. $\Delta H_{0}$(T) is measured at low
fields. The anisotropy of $\Delta H_{0}(T)$ is small
[Fig.~\ref{fig:linewidth}(b)]. $\Delta H_{0}(T)$ increases smoothly with
decreasing temperature from 300~K to about 75~K and decreases steeply
below 50~K in all directions.

In contrast, the linewidth due to fluctuations of the field-induced
staggered magnetization, $\Delta H_{\mathrm{fl}}(\mathbf{H},T)$ rapidly
\emph{increases} at high fields as $T_N$ is approached. At fixed
temperatures near $T_N$, the increase of $\Delta
H_{\mathrm{fl}}(\mathbf{H},T)$ with $\mathbf{H}$ is stronger than linear
[Fig.~\ref{fig:linewidth}(a)]. $\Delta H_{\mathrm{fl}}(\mathbf{H},T)$ is
largest in the plane perpendicular to $\mathbf{D}$, as shown in
Fig.~\ref{fig:linewidth}(b) for $\mathbf{H}\parallel\mathbf{c}$ and
$\mathbf{H}\parallel\mathbf{c}\times\mathbf{D}_A$. On the other hand,
there is no field-induced line broadening parallel to $\mathbf{D}_B$;
$\Delta H$ is field independent above 30~K
[Fig.~\ref{fig:linewidth}(b)].

Like the $g$~shift, the field-induced linewidth of individual ET layers
is independent of the magnetization in adjacent layers. $\Delta
H_{\mathrm{fl}}(\mathbf{H},T)$ is isotropic in the plane perpendicular
to the DM vector: it is about the same in the $\mathbf{c}$ direction
where $\mathbf{H}\perp \mathbf{D}$ in all layers, and in the
$\mathbf{c}\times\mathbf{D}_A$ direction where $\mathbf{H}\perp
\mathbf{D}$ in every second layer. In the $\mathbf{b}$ direction the
induced linewidth is smaller than the linewidth induced by fields of the
same magnitude along $\mathbf{c}\times\mathbf{D}_A$. It appears from
Fig.~\ref{fig:linewidth}(c) that within experimental precision only the
component of $\textbf{H}$ along $\mathbf{c}\times\mathbf{D}_A$ broadens
the line. This also supports that the magnetization is induced
independently in adjacent layers since $\mathbf{M}_s $ is the same in
all layers for $\mathbf{H}\parallel\mathbf{b}$.

It is natural to assume that $\Delta H_{\mathrm{fl}}(\mathbf{H},T)$ is
dominated by fluctuations of $\mathbf{M}_s$. Although in general, both
longitudinal and transverse fluctuations with respect to the external
field increase the linewidth, near $T_N$ $\mathbf{M}_s$ is much larger
than the longitudinal magnetization.

We know of no calculations of the ESR linewidth for the 2D anisotropic
Heisenberg antiferromagnet. On the other hand, the excitations and the
ESR spectrum of the 1D chain were extensively studied
\cite{Oshikawa1997, Affleck1999, Dender1997, Zvyagin2005}. The DM
interaction in an external magnetic field acts approximately as a
staggered magnetic field, $\mathbf{h}=(\mathbf{D}_{12}/J)\times
\mathbf{H}$ with an alternating sign for subsequent molecules along the
chain~\cite{Oshikawa1997}. (There is a further small contribution to
$\mathbf{h}$ from the in-plane alternating $g$ factor anisotropy).
$\mathbf{H}$ induces an anisotropic gap in the excitation spectrum
proportional to $(DH)^{1/2}$ for small $H$. The gap is largest for
$\mathbf{H}\perp\mathbf{D}$ and vanishes for $\mathbf{H} \parallel
\mathbf{D}$. At low temperatures $\mathbf{D}$ shifts the $q=0$ ESR mode.
Fluctuations of the staggered magnetization at the Larmor frequency at
finite wavelengths, $q$, broaden the ESR. The line broadening is
anisotropic and is roughly proportional to $h^2/T^2$ but absent if
$\mathbf{H}\parallel\mathbf{D}$.

The ESR in \mbox{$\kappa$-ET$_2$-Cl} resembles the ESR in 1D Heisenberg
chains with a DM interaction \cite{Zvyagin2005}, the main difference is
that \mbox{$\kappa$-ET$_2$-Cl} orders at a finite temperature. The
similarity with the 1D chain arises from the structure: the DM
interaction can be replaced by an effective staggered field,
$\mathbf{h}$ that is uniform within lines of dimers along $\mathbf{a}$
but alternates along $\mathbf{c}$ [Fig.~\ref{fig:lstruct}(b)]. In the
mean field approximation the frequencies of the two $q=0$ modes
are~\cite{Fouet2004}:
\begin{eqnarray}
   \hbar\omega _- &=& (hJ\mu_s)^{1/2} \label{eq1} \\
   \hbar\omega _+ &=& g_0\mu_BH\left(1 + \frac{Jh\mu_s}{2(g_0\mu_B)^2H^2}\right) \label{eq2}
\end{eqnarray}
to smallest order in $h$.

The ESR frequency above $T_N$ is given by $\omega _+$. For
$\mathbf{H}\perp\textbf{D}$ we have $h=HD_{12}/J$ and the magnetic field
dependent shift is:
  \begin{eqnarray}
  \Delta g=g_{0}(DM_{0}/8H) \mu_{s}/\mu_{B} \label{eq3}
  \end{eqnarray}
The measured $g$ shifts are in excellent agreement with Eq.~(\ref{eq3}).
Above $T_N$, the staggered moment $\mu_{s}$ decreases rapidly with
temperature and increases with the component of $\textbf{H}$
perpendicular to $\textbf{D}$. Moreover, $\mu_s$ in similar fields,
measured from $\Delta g$ and from the NMR Knight shift~\cite{Kagawa2008}
agree well (Fig.~\ref{fig:gfactor}).

We suggest that the magnetic field dependent NMR shift \cite{Kagawa2008}
and relaxation rate \cite{Hamad2005} in \mbox{$\kappa$-ET$_2$-Cl} are
also ``2D effects'' in the sense that they do not depend on interlayer
coupling. For $\mathbf{H}$ along $\varphi_{ab}=45^{\circ}$ the ESR
spectra of $A$ and $B$ layers are alternately affected and unaffected by
the field and we predict that this is the case for the NMR Knight shift
and relaxation also. We propose that the gap in the excitation spectrum,
$\hbar\omega _-$ is the reason for the anomalous increase of $T^{\ast}$
with field. Close to $T_N$, until $\omega_-$ is larger than the NMR
frequency $\omega_n$, fluctuations enhancing the nuclear spin
relaxation are suppressed. Above $T^{\ast}$ the gap is smaller than
$\omega_n$ and the fluctuations are restored.

The magnetic-field-induced increase of the static staggered
magnetization and magnetic fluctuations in layers with
$\mathbf{H}\perp\mathbf{D}$ is undoubtedly a purely 2D effect depending
only on intralayer interactions. In layers with
$\mathbf{H}\parallel\mathbf{D}$, both the static component of
$\mathbf{M}_s$ and its fluctuations are insensitive to the large
increase of $\mathbf{M}_s$ in adjacent layers, even at temperatures very
close to $T_N$ (Figs.~\ref{fig:gfactor} and \ref{fig:linewidth}). The
absence of correlation between the magnetization of adjacent layers
(measured in magnetic fields) suggests that interlayer correlations are
unimportant in determining the phase transition temperature $T_N$ (in
$H=0$); and in this sense \mbox{$\kappa$-ET$_2$-Cl} is a 2D magnet.

Finally we discuss the dimensionality of the phase transition. The
absence of correlation between the magnetic order of adjacent layers
raises the question whether the phase transition at $T_N=23\rm\,K$ in
the absence of $H$ is driven by the slightly anisotropic intralayer
exchange alone, or does the interlayer coupling enforce order in the
third direction? Below $T_N$ the weak ferromagnetism of isolated layers
have a twofold degeneracy in $H=0$. The full 3D ordering (i.e., along
the third direction) under cooling in $H=0$ below $T_N$ is an open
question. It may be ferromagnetic, antiferromagnetic, and it may
happen at any temperature below $T_N$, but we believe it must be very
sensitive to crystalline imperfections. In our experiments under
magnetic fields the degeneracy is lifted even in the
$\mathbf{H}\parallel\mathbf{D}$ layers by misalignment of the field, the
$g$ factor anisotropy between layers and other small anisotropic
effects. Thus, as the temperature is lowered there is a crossover to a
weak ferromagnetic phase. In the experiment of
Ref.~\onlinecite{Kagawa2008} the ferromagnetic order was enforced at low
temperatures by a magnetic field prior to the determination of $T_N$ in
$H=0$.

We are indebted to F.~Mila and K.~Penc for stimulating discussions and
acknowledge the Hungarian National Research Fund OTKA NN76727, CNK80991,
CK84324, K107228, the New Hungary Development Plan
T\'AMOP-4.2.2/B-10/1--2010-0009, and the Swiss NSF and its NCCR
``MaNEP''.

%\bibliography{etflukt73}

%

\end{document}